# Eshelby-twisted 3D moiré superlattices

Zhigang Song[1], Xiaotian Sun[2], Lin-Wang Wang[1]*

1 *Computational Research Division, Lawrence Berkeley National Laboratory, Berkeley, CA 94720, USA*
2 *College of Chemistry and Chemical Engineering, and Henan Key Laboratory of Function-Oriented Porous Materials, Luoyang Normal University, Luoyang 471934, P. R. China*

Corresponding author: Linwang Wang
Email: lwwang@lbl.gov

**Abstract**

Twisted bilayers of van der Waals materials have recently attracted great attention due to their tunable strongly correlated phenomena. Here, we investigate the chirality-specific physics in 3D moiré superlattices induced by Eshelby twist. Our direct DFT calculations reveal helical rotation leads to optical circular dichroism, and chirality-specific nonlinear Hall effect, even though there is no magnetization or magnetic field. Both these phenomena can be reversed by changing the structural chirality. This provides a way to constructing chirality-specific materials.

# I. INTRODUCTION

Due to weak van der Waals (vdW) interaction, a bilayer or few-layers of two-dimensional (2D) materials can be twisted by any angle, resulting in quasicrystals or 2D superlattices called moiré superlattices.[1] The interlayer interaction stemmed from moiré superlattices enforces a quantum confinement on each separate layer, flattening the bands near the Fermi level. The twisted layers lead to a series of interesting tunable phenomena, such as nontrivially topological phases, superconductivity and Mott insulators.[2-7] So much so the twisted homo and hetero bilayers gives rise to a new field called twistronics.[8,9] Until now, most of the experimental and theoretical endeavors are focused on two-dimensional structures including bilayers, or trilayers of graphene and transition metal dichalcogenide.[10-15] The electronic structure on the 2D plane are well studied, but no energy dispersion or electron conductivity in the vertical direction are involved in. For a long time, moiré patterns are thought as a concept only in the field of 2D materials.

Mathematically, it is possible to twist vdW materials layer by layer with certain angles (Eshelby twist), forming a 3D superlattice with periodicities in all three directions (illustrated in Fig. 1a). Such 3D helical crystals have structural chirality. Chiral phenomena are an intensely studied topic in the field of condensed matters, such as chiral bands in Weyl semimetals,[16] polarization of valley degree of freedoms[17] and chiral pairing in superconductivity[18]. Structural chirality will play a similar and important role as in the electronic chiral structures,[19] giving rise to pseudomagentic field[20] and nonreciprocal effects[21,22]. Eshelby twist can combine the features of 2D moiré superlattices and chirality in 3D to realize fantastic physics. More importantly, the chirality can be designed by controlling the twist angle. Due to the large computational cost in density functional theory (DFT) calculations and the challenges in building an effective Hamiltonian, the electronic structures of 3D moiré superlattices have rarely been studied till now, although they are highly expected.[23] Besides, recent experiments show helically twisted structures can be synthesized using the method of vapor-liquid-solid growth.[24,25] Thus, it is urgent to explore the functionalities and potential applications of these artificially-constructed structures to guide the future experiments. We apply our large-scale DFT methods and codes[26] in Eshelby-twisted structures to predict a series of general phenomena, which are not material-specific.

We take α-GeTe as an example to discuss the basic geometry of gradually twisted vdW materials and the significant chirality-dependence of the optical and transport properties. The chiral structures can lead to chirality-specific nonlinear Hall effect and prefer to absorb left-handed (or right-handed) circularly polarized photons. When the structural chirality is changed, both the Hall current and the helicity of optical absorption are reversed. Usually, a prerequisite for the presence of the conventional Hall or anomalous Hall effects is the breaking of time-reversal symmetry by either a magnetic field or intrinsic magnetization.[27,28] Here, the time-reversal symmetry is intact, and an additional vertical electric field is applied in place of the magnetic field. In applications, it is much easier to apply a local electric field in the nanoscale than a magnetic field. Thus, such helical 3D moiré superlattices can be used to fabricate nano-devices fully controlled by electric fields in the microelectronics.[29]

## II. ELECTRONIC STRUCTURES AND CHIRAL RESPONSES

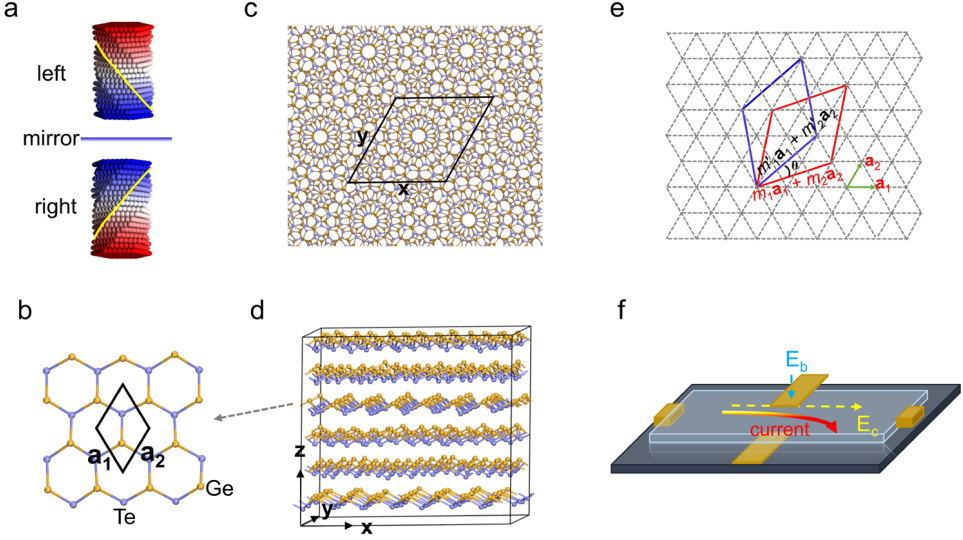

Figure 1. (a) Illustration of twisted layers and their mirror imagine. (b) Unit cell of the primary layers of α-GeTe. (c) Top and (d) side views of the three-dimensional moiré superlattice with an interlayer twist angle of approximately 20° after DFT relaxation. The frame of solid lines illustrates the unit cell of moiré superlattices. (e) Schematic relationship between 2D moiré superlattices and twist angle. (f) Illustration of nonlinear Hall effect. Golden cuboids are the electrodes to generate the electric fields $E_b$ and $E_c$.

We will first discuss how to construct a 3D moiré superlattice based on the illustration in Fig. 1a. Only some special angles $\theta$ generate structures with periodicity in all three dimensions. First, if the original layers hold a $C_n$ axis perpendicular to the materials plane, it is expected that $\theta=360°m/(nj)$ can lead to a z-direction periodically stack, where $j$, $m$ and $n$ are positive integers. The z-direction (perpendicular to the 2D materials) periodicity is $j$ atomic layers. Second, to form a 3D periodic structure, any two adjacent layers must form a commensurate 2D moiré supperlattice. The conditions for rectangular and triangle systems to form 2D commensurate superlattices are summarized in previous review work.[1] Let's use $\mathbf{a}_1$ and $\mathbf{a}_2$ to denote the original primary lattices in a single layer (Fig. 1b). Here, $a_1=a_2=4.23$Å, and $a_3=4.45$Å is the averaging distance between two layers for α-GeTe. To construct a commensurate moiré superlattice, one needs to find integer pairs $(m_1, m_2)$ and $(m'_1, m'_2)$ satisfying $|m_1\mathbf{a}_1 + m_2\mathbf{a}_2| = |m'_1\mathbf{a}_1 + m'_2\mathbf{a}_2|$, and the rotation angle $\theta$ is the tilting between $m_1\mathbf{a}_1 + m_2\mathbf{a}_2$ and $m'_1\mathbf{a}_1 + m'_2\mathbf{a}_2$. The supercell has a same shape as the primary cell of original layers, and thus one can find the other lattice vector by rotating $m_1\mathbf{a}_1 + m_2\mathbf{a}_2$ with 60° (90°) in triangle (square or rectangle) systems. In case of $|\mathbf{a}_1|=|\mathbf{a}_2|$, one possible choice is $m'_1= m_2$ and $m'_2= m_1$. In a triangular lattice, for example α-GeTe, , the twist angle can be determined as $\theta = \arccos(\dfrac{m_1^2 + 4m_1m_2 + m_2^2}{2(m_1^2 + m_1m_2 + m_2^2)})$ based on Fig. 1e.[30] In a square lattice, $\theta = \arccos(\frac{2m_1 m_2}{m_1^2 + m_2^2})$. In rectangular lattices, we can have: $m'_1 = m_1$, $m'_2 = -m_2$, and $\theta = \arccos(\frac{m_1^2 a_1^2 - m_2^2 a_2^2}{m_1^2 a_1^2 + m_2^2 a_2^2})$. To further rotate the system in the third layer, we can view the constructed two-layer system having the same square or triangle primary cell (as the original one) with $m_1\mathbf{a}_1 + m_2\mathbf{a}_2$ as one of its primary lattice (another primary lattice can be obtained by 90° or 60° rotation from $m_1\mathbf{a}_1 + m_2\mathbf{a}_2$ for square and triangle lattice, respectively). Then we can repeat the above procedure to rotate the same $\theta$ angle. If $\theta$ satisfies the $\theta=360°m/(nj)$ rule, then after the $j_{th}$ rotation, we come back to the original lattice orientation. For example, the original two-dimensional layers have $C_3$ symmetry, $n$ is 3. If the twist angle is 48°, the period in the z-direction is 5 and $m$ is 2.

Although the above procedure can produce a 3D moiré superlattice with the exactly same $\theta$ rotation each layer, the resulting periodicity in the x-y plane can be extremely large for density functional theory (DFT) calculations. (It applies the $\mathbf{a}_1' = m_1\mathbf{a}_1 + m_2\mathbf{a}_2$ operation $j$ times). Atomic number in a cell of 3D moiré superlattices increases rapidly as the twist angle decreases. Although each unit cell of the original GeTe layers include only one Ge and one Te atom (see Fig. 1b), the total number of atoms is still too large.

In reality, we will relax some requirements, e.g., the above requirements are satisfied approximately in order to yield a smaller 3D moiré superlattice. For example, $\theta=360°m/(nj)$ will only be satisfied, and the $\theta$ for each layer can be slightly different. To obtain the electronic structures of the 3D moiré superlattices, we build three structures with $\theta \approx 30°, 20°, 15°$, which has relatively small lattices of $|m_1\mathbf{a}_1 + m_2\mathbf{a}_2|$. In fact, the periodicity is absent in the lateral direction, forming a quasicrystal, if $\theta$ is exactly 30°, 20° or 15°. In the first structure with approximate twist angle of 30°, which has 4 vertical layers in each z-direction period, the actual twist angle series here are 0°, 27.51°, 60°, 87.51° for different layers, instead of 0°, 30°, 60°, 90°. In the second structure with approximate twist angle of 20°, which has 6 vertical layers in each z-direction period, the actual twist angle series are 0°, 21.80°, 38.21°, 60°, 81.80°, 98.21° for different layers instead of 0°, 20°, 40°, 60°, 80°, 100°. In the third structure with gradual twist angle of 15°, the actual twist angle series are 0°, 15.52°, 27.51°, 38.21°, 60°, 75.52°, 87.51°, 98.21° for different layers instead of 0°, 15°, 30°, 45°, 60°, 75°, 90°, 105°. The three unit cells of 3D moiré superlattices are $\sqrt{13}a_1 \times \sqrt{13}a_2 \times 4a_3$, $7a_1 \times 7a_2 \times 6a_3$ and $\sqrt{559}a_1 \times \sqrt{559}a_2 \times 8a_3$. There are 4, 6 and 8 original layers in 3 supercells with twist angle of 30°, 20° and 15°, respectively. More exactly, their z-direction periodicities are 17.83, 26.74 and 35.65 Å. Even so, the total number of atoms in each supercell is still as large as 104, 588 and 8944, respectively.

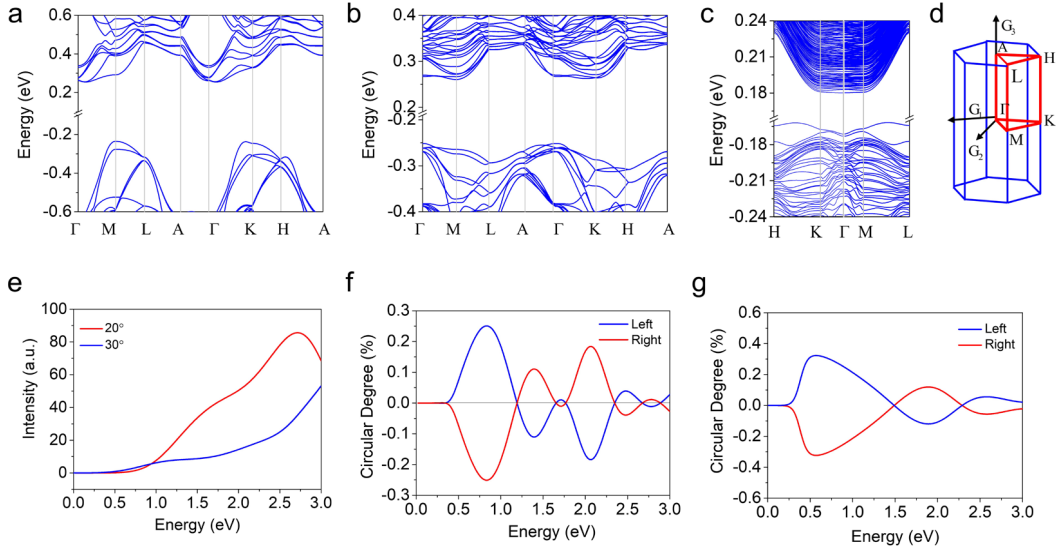

Figure 2. Energy dispersion along high-symmetric path in the reciprocal space. The twist angle is (a) 30°, (b) 20° and (c) 15°. (d) Brillouin zone and high-symmetric points. (e) Average absorption intensity as a function of excitation energy. (f) and (g) Comparison of circularly polarized degree of optical absorption between left- and right-handed structures: (f) $\theta\approx\pm30°$ and (g) $\theta\approx\pm20°$. Gaussian broadening of 0.05eV is used to smear out the energy conservation in the delta function

Note, if one finds a solution for $\theta$ ($\theta<60°$) in whatever method, the $-\theta$ is also a solution. These two structures will have opposite chirality, as illustrated in Fig. 1a. We define the left (right) handedness, when $\theta$ is negative (positive). Using DFT calculation, the atomic positions are fully optimized in all structures. As an example, atomic structures with $\theta\approx20°$ is shown in Fig. 1c&d. Compared with a single free-standing layer, the atoms in 3D moiré superlattices slightly rearrange themselves after optimization.

The band structure is the same in both left-handed and right-handed structures. The calculated band structures under different twist angles are shown in Fig. 2a-c, and the corresponding high-symmetric path is shown in Fig. 2d. The monolayer GeTe has a band gap of 0.723 eV. The band gaps are 0.489, 0.512 and 0.34 eV in 3D moiré superlattices for $\theta\approx30°$, 20° and 15°. Due to the atomic reconstruction and interaction in the vertical direction, the band gap has a non-monotonic relationship with the twist angle $\theta$. Band gap becomes significantly smaller at 15° probably due to the state localization in this large system. The energy dispersion vertical to the material planes, such as M-L and K-H, cannot be neglected compared with the in-plane dispersion. The three structures have very different band structures, showing a large space for band-structure engineering. After the atomic relaxation (see Fig. 1c&d), the interlayer interaction modifies the atomic positions, resulting in layer distortion compared with a free-standing layer. Some symmetry-protected degenerations are lifted by the distortion. The band structures are not simply folded from the original band structures of primary layers. The nearly degenerate bands will induce large Berry curvature and further large Berry curvature dipole.[20]

The calculated charge density of valence band maxima is shown in Fig. 3 when $\theta\approx20°$. The slides of charge density at different heights show the evolution in the z-direction. The charge density is also helical in the z-direction.

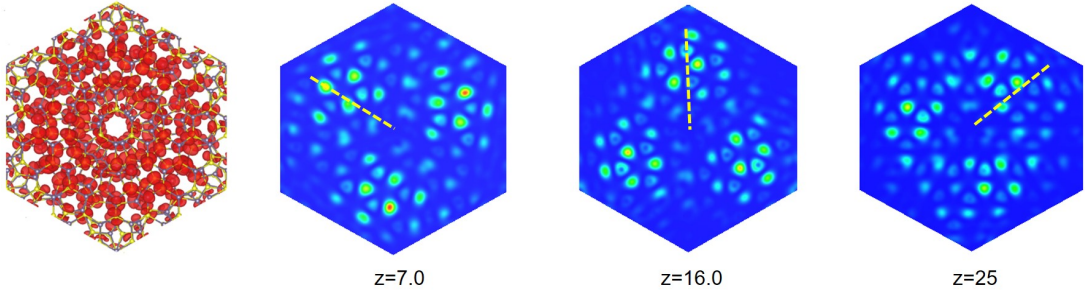

z=7.0   z=16.0   z=25

Figure 3. Charge density of valence band maxima in the structure with twist angle of 20°. The left one is the top view of total charge density. Right three are the slides at different heights z. Yellow dash lines are guides for the eyes

## III. CHIRALITY-SPECIFIC CIRCULAR CHRISOM

Owing to the difference of the electronegativity of Ge and Te atoms, there is a local electric dipole between the two bonded Ge and Te atoms. The interlayer electric dipoles show a series of polar vortices, which can be described by an order parameter $\mathbf{T} = \mathbf{r} \times \mathbf{P}$. Here $\mathbf{r}$ and $\mathbf{P}$ are position (with $\mathbf{r}$ having its origin defined at the high-symmetric points) and electrical dipole, respectively. Following the previous work,[31] the chirality ($C \backsim \mathbf{T}\cdot\hat{z}$) of the screw-type structure is determined by the ferro-rotational moment $\mathbf{T}$ combined with the $\hat{z}$ vector representing the structural rotation direction. When $C$ is positive, we have right handness, otherwise, you have left handness. When the rotation become $-\theta$, the sign of $C$ will also change. The chirality of 3D moiré superlattices and the chirality manifest themselves by the optical circular dichroism and chirality-specific nonlinear Hall effect.[31] When the structural chirality is changed, the chirality of absorbed photons and nonlinear Hall current are also reversed.

The circularly polarized dichroism of optical absorption is defined as[32-34]

$$\eta(\Delta E) = \frac{I_+(\Delta E) - I_-(\Delta E)}{I_+(\Delta E) + I_-(\Delta E)} \times 100\% \qquad (1)$$

where $I_+$ and $I_-$ are the absorption strength for left- and right-handed circularly polarized light. The absorption probability of left- (right-) handed light is calculated as following

$$I_{+(-)}(\Delta E) = \frac{2\pi\hbar^3 e^2}{\Delta E^2 m_e^2 c^2}|\tilde{\varepsilon}|^2 \sum_{m,n}\int d\mathbf{k}\,|\langle\varphi_{m,\mathbf{k}}|p_x \pm ip_y|\varphi_{n,\mathbf{k}+2\pi\hat{z}/\lambda}\rangle|^2\,\delta(E_{n,\mathbf{k}+2\pi\hat{z}/\lambda} - E_{m,\mathbf{k}} - \Delta E) \qquad (2)$$

where $m$ and $n$ indexes the occupied and unoccupied band states, respectively. $p_x$ and $p_y$ are the normal momentum projection in the $x$- and $y$-direction. $\lambda$ is the wave length of incident light, and $\Delta E$ is the excitation energy. $\tilde{\varepsilon}$ is the electric field of the light. $\hat{z}$ is a unit vector parallel to the rotation axis. $\mathbf{k}$ is k-point of the supercell. $m_e$, $c$ and $\hbar$ are the electron mass, light velocity and reduced Planck's constant.

The optical absorption is nonzero only when the momentum in the initial state and finial states is slightly different. Usually, photon momentum is assumed as zero due to its small magnitude. Here, it is critical to include the 2πz/λ term here in order to have an effect for the light propagation direction. This also shows, the maximum circular dichroism might be reached when the periodicity of the helix structure is in the same order of the light wavelength. Due to the limitation of our calculation power, we cannot reach such long periodicity, but one can predict that much larger circular dichroism values can be achieved by using smaller $\theta$ values. If the incident light is in the direction perpendicular to the layers, the average optical absorption $(I_+ + I_-)/2$ is shown in Fig. 2e. The calculated results of circularly polarized degree are shown in Fig. 2f&g. The circularly polarized degree $\eta(\Delta E)$ of left- and right-handed 3D moiré superlattices has a same amplitude but opposite sign, as dictated by the chiral symmetry of the structures. The dichroism oscillates in sign with the incoming photon energy. The oscillation is slower for the larger system (20° structure). The average amplitude of circular polarization roughly increases as the decreasing twist angle. As stated above, the amplitude of circular polarization reaches its maxima, when the helical period in the z-direction matches the wavelength. In natural inorganic crystals, the circular dichroism usually are caused not by structural helicity, instead are caused by magnetic field or internal magnetization due to the break of time inversion symmetry.[35,36]

## IV. CHIRALITY-DEPENDENT NONLINEAR HALL EFFECT

Similar to optical circular dichroism, the structural helicity will induce a chirality-specific nonlinear Hall effect. When the current flows across the helical vortex in one direction, (for example x-direction), under a vertical electric field (z-direction), the current will be bended in the third direction (y-direction). This effect can be calculated using the Berry curvature dipole ($\Lambda$). Berry curvature dipole is a tensor in 3D materials and defined as [37]

$$\Lambda_{ab} = \sum_n \int \frac{\partial \Omega_{nb}(\mathbf{k})}{\partial k_a} f_0 d\mathbf{k} = \sum_n \int D_{ab}(\mathbf{k})\delta(E_k - \mu)d\mathbf{k} \qquad (3)$$

where $\Omega_{nb}$ is the Berry curvature of the $n_{th}$ band projected on $b$ direction. $f_0$ is the Fermi-Dirac occupation function, $\mu$ is the chemical potential. $D_{ab}(\mathbf{k}) = \Omega_{nb}(\mathbf{k})v_a(\mathbf{k})$ is Berry curvature dipole density and $v_a(\mathbf{k})$ is the electron group velocity. Within the Boltzmann picture of transport and the relaxation time approximation, Inti Sodemann et al. derived the Hall-current response to the electric field up to the second order.[37] Nonlinear Hall effect in inversion-asymmetric systems has attracted great attention in recent years.[38-44] Phenomenologically, the nonlinear Hall current can be expressed as: $j_a = \{\chi_{abc}\tilde{\varepsilon}_b\tilde{\varepsilon}_c^* + \chi_{abc}\tilde{\varepsilon}_b\tilde{\varepsilon}_c e^{2i\omega t}\}$, and $\tilde{\varepsilon} = \mathrm{Re}\{\tilde{\varepsilon}_0 e^{i\omega t}\}$ ( the repeated subscripts indicated contraction through summation of that index, same convention below). The coefficient $\chi_{abc}$ is related to Berry curvature dipole as: $\chi_{abc} = -\varepsilon_{adc}\frac{e^3\tau}{2\hbar^2(1+i\omega\tau)}\Lambda_{bd}$,[45] where $\tau$ is electron relaxation time. $\varepsilon_{abd}$ is Levi-Civita anti-symmetric tensor. $\varepsilon_{abd}$ is 1, after an even permutation of any two indexes, -1 after an odd permutation, and 0 if any index is repeated. If the electric field is a dc field, the frequency is zero $\omega \to 0$, then $\chi_{abc} = -\varepsilon_{adc}\frac{e^3\tau}{2\hbar^2}\Lambda_{bd}$. It is interesting that some components of Berry curvature dipole tensor change signs when the chirality of the structure changes from left-handedness into right-handedness. The other components are independent of the chirality. A left-handed structure is the mirror imagine of the right-handed one. According to the mirror reflection, we can divide the Berry curvature dipole into symmetric ($\Lambda^+$) and anti-symmetric part ($\Lambda^-$). $\Lambda^{\pm} = \frac{1}{2}(\Lambda \pm M\Lambda M^{-1})$, where $M$ is the mirror operator. Without losing generality, we can assume the mirror plane is the x-y plane (illustrated in Fig. 1a). $D_{xx}(\mathbf{k})$, $D_{yy}(\mathbf{k})$, $D_{zz}(\mathbf{k})$, $D_{xy}(\mathbf{k})$, $D_{yx}(\mathbf{k})$ belong to $\Lambda^-$. $D_{yz}(\mathbf{k})$, $D_{zy}(\mathbf{k})$, $D_{xz}(\mathbf{k})$, $D_{zx}(\mathbf{k})$ belong to $\Lambda^+$. The DFT calculated results of valence band of the $\theta \approx 30°$ structure are shown in Fig. 4. The structures of $\theta \approx 20°$ exhibit similar dependence of chirality to the results of $\theta \approx 30°$. The calculated Berry curvature dipole as a function of energy (or chemical potential $\mu$) near the valence band maximum is shown in Fig. 5. According to Eq. (3), Berry curvature dipole as a function of energy can be used to calculate the corresponding nonlinear Hall conductance under different doping level. As we can see, we the increase of doping level, the nonlinear Hall conductance will be increased significantly.

If we apply two electric fields, one in x-y plane, another in z-direction, the current will be bended inside the plane, resulting in a Hall-like effect. More specifically, the Hall current is $j_x = \frac{e^3\tau}{2\hbar^2}(\Lambda_{zz} - \Lambda_{yy})$, if the electric fields are applied in y- and z-directions. Similarly, $j_y = -\frac{e^3\tau}{2\hbar^2}(\Lambda_{zz} - \Lambda_{xx})$ if the electric fields are applied in x- and z-direction. Note, here, the electric current is the bulk total current in all layers, not in individual layers. According to Fig. 4 and Fig. 5, $D_{zz}(\mathbf{k})$ is much smaller than $D_{yy}(\mathbf{k})$ and $D_{xx}(\mathbf{k})$, thus $j_x$ (or $j_y$) is dominated by $D_{yy}(\mathbf{k})$ and $D_{xx}(\mathbf{k})$. If the twist angle is opposite, the signs of $D_{xx}(\mathbf{k})$, $D_{yy}(\mathbf{k})$ and $D_{zz}(\mathbf{k})$ are all reversed.

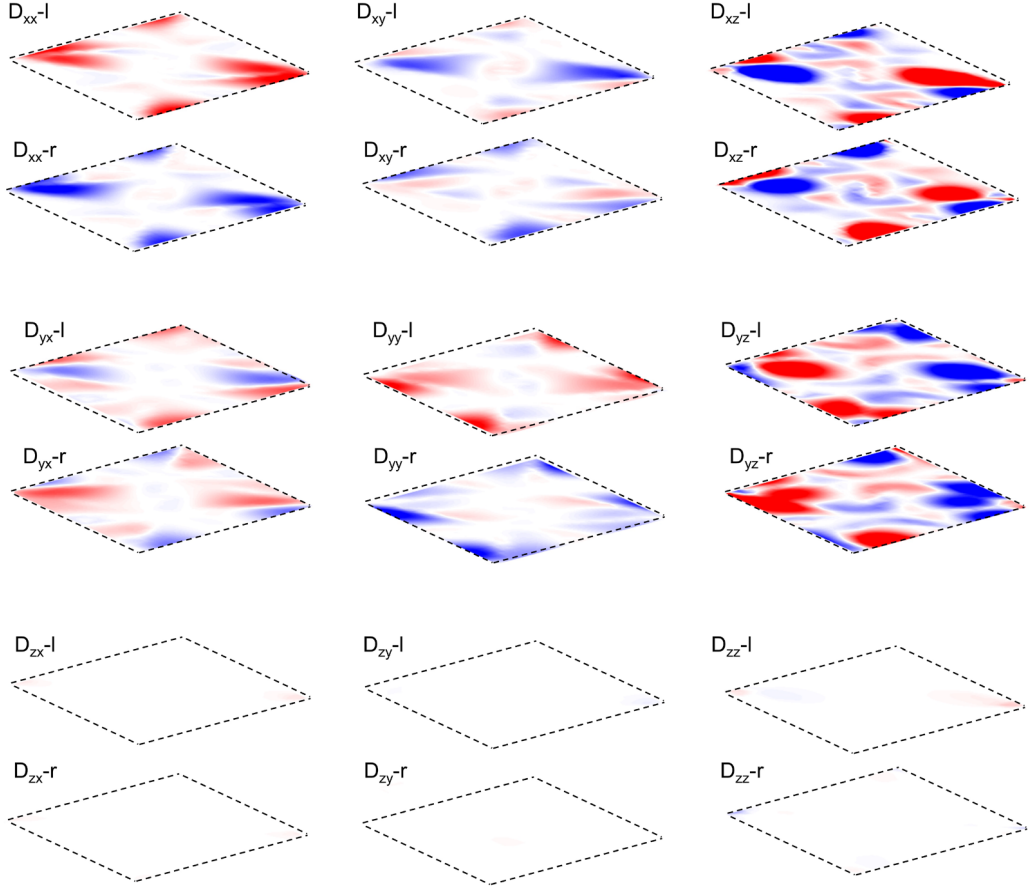

Figure 4. Momentum-resolved Berry curvature dipole tensor of the valence bands in the reciprocal space. $k_x$-$k_y$ planes are slides where $k_z$ is 0. *l* and *r* represent the results in the left-handed and right-handed structures, respectively. Twist angle is $\theta\approx30°$.

This phenomenon seems like the conventional and anomalous Hall effect. Remarkably, different from Hall effect, which origin from the Lorentz force, and anomalous Hall effect, which stems from the nonzero integration of Berry curvature on the Fermi surface,[28] here both Lorentz force and Berry curvature integration are zero. This nonlinear Hall effect is induced purely by a geometric effect, which are related to nonvanishing Berry curvature dipole on the Fermi surface. Usually, the spin orbital coupling (SOC) is critical to anomalous Hall effect. However, the SOC does not play a significant role in the case of nonlinear Hall effect. For example, the Berry curvature of Eq. (3) can be calculated without SOC. It is worthy to mention that the calculated Berry curvature dipole is in the same order as the value in 1-T' $WTe_2$, where the phenomena of Berry curvature dipole have been observed experimentally.[41,46] We believe the nonlinear Hall effect in 3D moiré supperlattices is large enough for the current experimental observation.

As discussion in previous part, chirality is the projection of the spin (or spin-like quantities) onto the momentum vector. For an electron wave packet, the Berry curvature is related to the self-rotation of wave packet in real space, and Berry curvature is a generalized spin.[35] The velocity is different from the momentum by only a constant. The Berry curvature dipole density of $D_{ab}(\mathbf{k}) = \Omega_{nb}(\mathbf{k})v_a(\mathbf{k})$ in Eq. (3) actually has an intrinsic chirality, especially when a=b. Thus, nonlinear Hall effect also has an intrinsic chirality, if two electric field are not parallel to each other. It is not a surprise that nonlinear Hall effect can be coupled to chiral structures. That is why only $D_{xx}(\mathbf{k})$, $D_{yy}(\mathbf{k})$ and $D_{zz}(\mathbf{k})$ (or $\Lambda_{xx}$, $\Lambda_{yy}$ and $\Lambda_{zz}$) are reversed when the structural chirality is changed.

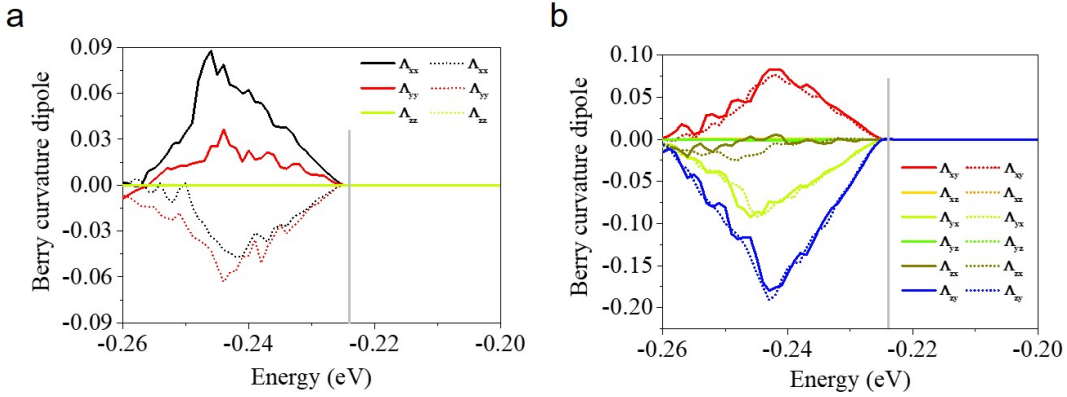

Figure 5.   Berry curvature dipole as a function of energy. Solid and dash curves represent the results in the left- handed and right-handed structures, respectively. Twist angle is $\theta \approx 30°$ here. (a) and (b) are the symmetric and antisymmetric parts of Berry curvature dipole tensor. The vertical and gray lines imply the valence band maximum.

## V. CALCULATION DETAILS

The calculation of Berry curvature dipole and circular dichroism is performed using the PWmat package[47] with ONCV norm conserving pseudopotenial, PBE exchange correlation function, and a plane wave energy cutoff of 50 Ry. When the spin orbital coupling (SOC) is turned on, the spin splitting is smaller than 10 meV. SOC only change slightly the values of the optical circular dichroism and Berry curvature dipole, making no major differences on the final conclusion. Thus, we neglected SOC during the calculation of electronic structures and Berry curvature. The atomic positions in the twisted structures are relaxed until the force on each atom is smaller than 0.01 eV/Å. The band structures and geometry optimization are calculated using code package of VNL ATK, and Single-Zeta basis set are used.

The Berry curvature is calculated as following:

$$\Omega_b^n(\mathbf{k}) = i\varepsilon_{cab}\sum_m \frac{\langle u_\mathbf{k}^n|\partial H/\partial_{k_a}|u_\mathbf{k}^m\rangle\langle u_\mathbf{k}^m|\partial H/\partial_{k_b}|u_\mathbf{k}^n\rangle - (a \leftrightarrow b)}{(E_\mathbf{k}^m - E_\mathbf{k}^n)^2} \quad (4)$$

where $m$ ($m \neq n$) runs over all the occupied bands and unoccupied bands. $u_\mathbf{k}^m$ is the nonperiodic part of Bloch states in the momentum space. In DFT calculation, $\langle u_\mathbf{k}^n|\partial H/\partial_{k_b}|u_\mathbf{k}^m\rangle$ can be obtained by post-processing. The **k**-resolved Berry curvature dipole density is calculated by $D_{ab}(\mathbf{k}) = \Omega_{nb}(\mathbf{k})v_a(\mathbf{k})$. To ensure the reliability, we compute the Berry curvature and Berry curvature dipole using a 21×21×1 **k**-mesh.

## ACKNOWLEDGEMENTS

This work was supported by the U.S. Department of Energy, Office of Science, Basic Energy Sciences, Materials Sciences and Engineering Division under Contract No. DE-AC02-05-CH11231 within the beyond-Moore's law LDRD project.